\documentclass[aps, prx, reprint]{revtex4-1} 
\usepackage{amssymb}  
\usepackage{amsfonts}  
\usepackage{amsmath}  
\usepackage{amsthm}  
\usepackage{bm}  
\usepackage{bbm}  
\usepackage[colorlinks=true]{hyperref}
\usepackage{graphicx}
\usepackage{tabularx}
\usepackage{bbding}

\begin{document}

\title{Charge-spin mutual entanglement: A case study by exact diagonalization of the one hole doped $t$-$J$ loop}
\author{Wayne Zheng}
\affiliation{Institute for Advanced Study, Tsinghua University, Beijing, 100084, China}
\author{Zheng-Yu Weng}
\affiliation{Institute for Advanced Study, Tsinghua University, Beijing, 100084, China}
\affiliation{Collaborative Innovation Center of Quantum Matter, Tsinghua University, Beijing, 100084, China}

\begin{abstract}

A doped Mott insulator exhibits peculiar properties associated with its singular sign structure. As a case study, we investigate the ground state and excitations of finite-size Heisenberg loops doped with one hole by exact diagonalization. We find that there appear a series of quantum critical points (QCPs), which separate regimes by distinct total momenta along the axis of the ratio $J/t$ ($J$ and $t$ denote the superexchange coupling and hopping integral, respectively). Each QCP involves a crystal momentum jump with level crossing or merging of lowest energy levels. In contrast to the conserved total momentum, however, a broad momentum distribution of \emph{individual} electrons is also found, indicating charge incoherence/translational symmetry breaking in violation of the one-to-one correspondence. Such a charge incoherence is further related to quantum fluctuations or the transverse part of ${\bf S}^2=3/4$ with $S^z=\pm 1/2$ in the one-hole ground state. Turning off the phase-string sign structure, by contrast, we show that the total momentum of the ground state reduces to null in the whole regime of $J/t$ with no more QCP or incoherence. We introduce the so-called charge-spin mutual entanglement to characterize these novel properties, with the entanglement spectrum providing additional information on the charge incoherence, which capture the nature of strong correlation due to the many-body quantum interference. 

\end{abstract}

\maketitle

\section{introduction}

The physics of doped Mott insulators is believed to be closely related to the mechanism of high temperature superconductivity in the cuprate \cite{anderson1987the}\cite{RevModPhys.78.17}. The $t$-$J$ model is one of the simplest models describing the doped Mott insulator with the double-occupancy of electrons being projected out in the hole-doped case. Earlier on, this projection, or the presence of the so-called \emph{upper} Hubbard band due to interaction, has been argued by Anderson\cite{PhysRevLett.64.1839} as responsible for producing an \emph{unrenormalizable quantum phase shift} each time a hole is doped into the lower Hubbard band, leading to a generic non-Fermi liquid behavior. Such an unrenormalizable phase shift effect has been later quantitatively identified as the \emph{phase-string} sign structure upon doping, based on the $t$-$J$ \cite{PhysRevLett.77.5102}\cite{PhysRevB.77.155102} or Hubbard model\cite{PhysRevB.90.165120}.

\begin{figure}
\centering
\includegraphics[width=0.3\textwidth]{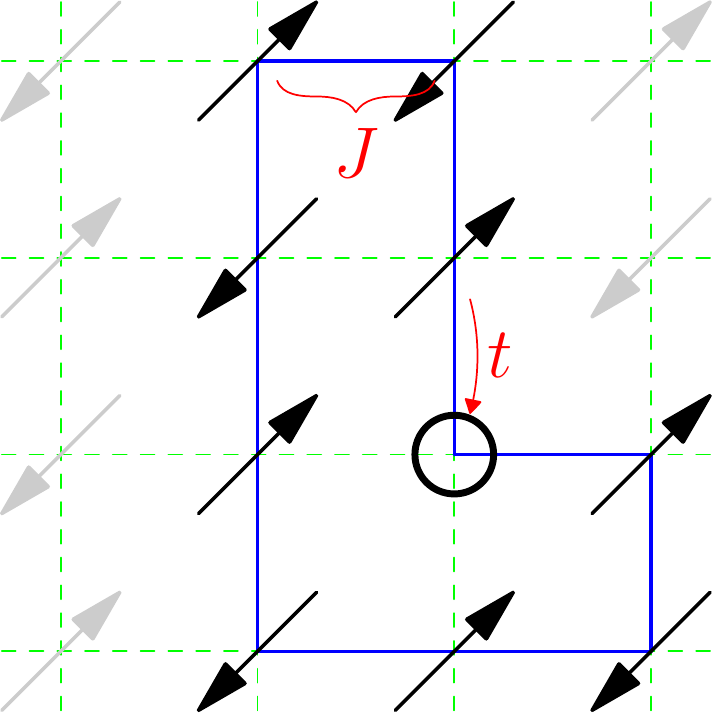}
\caption{A closed one-hole loop may be regarded as a subsystem of the 2D $t$-$J$ model in which the hopping and superexchange processes connecting to the rest of the 2D lattice are cut-off. The effect of the hidden sign structure in this specific case (cf. Appendix A) can be precisely studied by exact diagonalization, which sheds light on a general situation like 2D involving more complicate closed paths.  }
\label{fig:illustration}
\end{figure}

A many-body quantum mechanics involving the nonlocal phase-string sign structure is conceivably non-perturbative\cite{zaanen2009mottness}. Its consequences can be manifested in various limiting cases as well. Although the two-dimensional (2D) case is more realistic, a one-dimensional (1D) closed path in which a hole is going through may represent a sub-unit that the nontrivial sign structure plays an indispensable role. In particular, a true 1D finite-loop, with all the hopping and superexchange processes away from the loop being cut-off, as illustrated in Fig. \ref{fig:illustration}, is probably the simplest nontrivial limit that can be precisely studied by exact diagonalization (ED) \cite{arpackpp}.

In this paper, we study the essential role of the quantum interference of the nonlocal phase shift experienced by the hole that circles around a finite-size 1D loop in Fig. \ref{fig:illustration}. By using numerical ED technique, the total momentum of the many-body ground state is found to take a series of values, as a function of $J/t$, to result in a series of quantum critical points (QCPs) where lowest energy levels merge/cross and the total momentum jumps. Here, the total momentum remains \emph{conserved} due to the translation symmetry, but the momentum distribution of the \emph{individual} electrons is shown to exhibit a broad feature involving all the crystal momenta allowed in the finite-size loop. 
In particular, we show that the phase-string sign structure is associated with the quantum fluctuation or the transverse part of ${\bf S}^2=3/4$ in the one-hole case with $S^z=\pm 1/2$. It is such quantum fluctuation, in addition to the longitudinal $S^z=\pm 1/2$ component, that carries away a spread of momenta to render the doped hole ``incoherent'' or the translation symmetry breaking for the doped charge. By contrast, such unconventional QCPs immediately disappear if the quantum interference is turned off, with the total momentum reducing to null throughout the whole $J/t$. The results will be presented in Sec. II.

The Landau's one-to-one correspondence between the total momentum and the momenta carried by quasiparticles breaks down here. To characterize such a strongly correlated many-body system, we introduce a new kind of entanglement to describe the mutual interplay between the doped charge and the spin degrees of freedom in Sec. III. The spin-charge mutual entanglement entropy (MEE) and the corresponding entanglement spectrum (MES) are investigated, which can reproduce the QCPs, measure the strength of spin-charge entanglement/separation, and reveal the incoherence of the charge degree of freedom. In other words, such a mutual entanglement description may provide the most relevant quantum information on the strong correlation of Mott physics, which may be applied to a more general case like the 2D case. 

Since the above numerical results and analytic analysis will be based on the sign structure of the $t$-$J$ model, in Appendix A, we briefly outline some basic rigorous results which are valid for the $t$-$J$ Hamiltonian on a bipartite lattice of any size and dimensions including the present 1D loop. In particular, the so-called $\sigma\cdot{t}$-$J$ model is presented in Eq. (\ref{stj}), in which the phase-string sign structure is turned off. The distinction between the  $t$-$J$ and $\sigma\cdot{t}$-$J$ models may be most clearly seen in the following exact expressions for their partition functions\cite{PhysRevB.77.155102}
\begin{equation}
Z_{t\text{-}J}=\sum_{c}(-1)^{N^{\downarrow}_h[c]}\mathcal{W}[c],
\label{Ztj}
\end{equation}
and
\begin{equation}
Z_{\sigma\cdot t\text{-}J}=\sum_{c}\mathcal{W}[c],
\label{Zstj}
\end{equation}
in which the positive weight $\mathcal{W}[c]\geq{0}$ is the same for both models as a function of $t$,$J$ and the temperature, and $c$ denotes all the closed paths of the hole and spins. Apparently the two models differ only by the phase-string sign structure $(-1)^{N^{\downarrow}_h[c]}=\pm 1$ in Eq. (\ref{Ztj}), which is dependent on the parity of $N^{\downarrow}_h[c]$ that counts the total number of exchanges between the hole and $\downarrow$-spins for a given loop $c$ (cf. Fig. \ref{fig:phasestring} in Appendix A). By a comparative ED study of the $t$-$J$ and $\sigma\cdot{t}$-$J$ loops with one hole, the critical role played by such a peculiar sign structure in a doped Mott insulator can be then explicitly revealed.

\begin{figure}[th]
\centering
\includegraphics[width=0.48\textwidth]{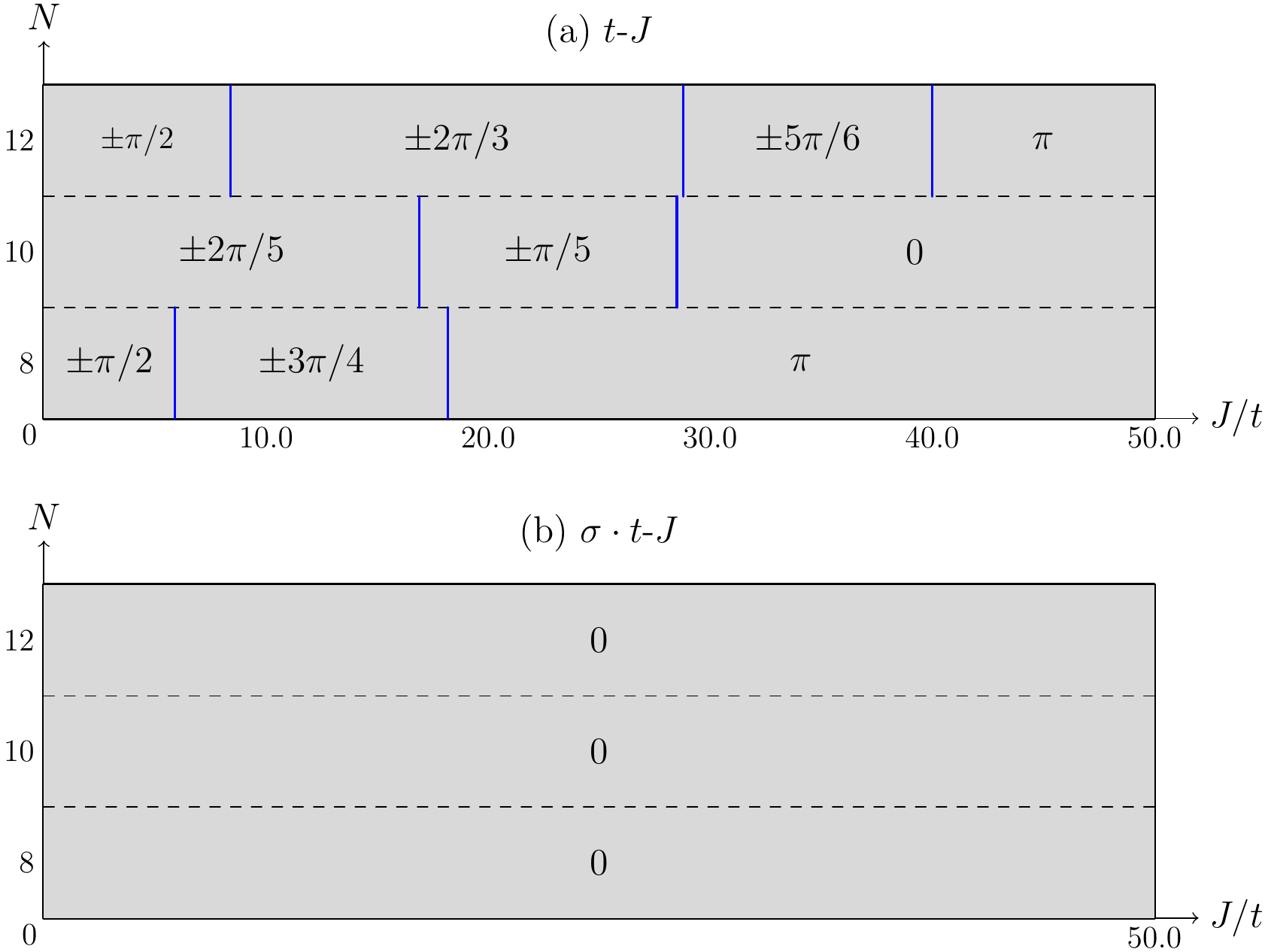}
\caption{The phase diagram of the one-hole loop with $N$ sites, characterized by a pair of total momenta $\pm K_0$ indicated in the boxes for the $t$-$J$ model (a) and $\sigma\cdot{t}$-$J$ model (b). Note that for each nontrivial regime in (a), which is characterized by a momentum $K_0$ other than $0$ and $\pi$, there is always another degenerate ground state with momentum $-K_0$ (cf. Fig. \ref{fig:energy_len-10}). Define $J_c$ to separate the trivial and nontrivial phases in the $t$-$J$ loop, and one has $J_c/t=18.2$, $28.5$, and $40.0$, corresponding to $N=8$, $10$, and $12$, respectively.}
\label{fig:crystal_momenta}
\end{figure}

\section{Basic results of one hole in the finite-size loop}

\subsection{Phase Diagram} 

Although a finite-size system is considered here, in order to distinguish distinct many-body ground states which cannot be analytically connected by tuning a parameter such as $J/t$, we shall still use the terminology ``phase diagram'' in a loose way. Namely, each ``phase'' will refer to a smooth regime of $J/t$ where no level crossing happens. In this sense, Fig. \ref{fig:crystal_momenta} (a) illustrates the ``phase diagram'' of one-hole loops with sizes $N=8$, $10$ and $12$ versus $J/t$ for the $t$-$J$ model under periodic boundary condition (PBC). 

As determined by ED, Fig. \ref{fig:crystal_momenta} (a) shows that the ground state exhibits a series of non-trivial total momenta $\pm K_0$'s with the presence of double degeneracy. Here the total crystal momentum is determined by diagonalizing the matrix $\langle\psi_{i}|T|\psi_{j}\rangle$ where $i, j=1, 2$ denotes the two degenerate ground states and $T$ the translational operator. A crystal momentum depends on the lattice size $N$ as $2n\pi/N, n=\pm 1, \pm 2\cdots$, and the jumps specify a series of QCPs in Fig. \ref{fig:crystal_momenta} (a). By contrast, the total momentum reduces to a trivial one at $0$ or $\pi$ beyond the largest critical $J_c/t$ (cf. the caption of Fig. \ref{fig:crystal_momenta})  where the ground state becomes non-degenerate.

\begin{figure}[]
\centering
\includegraphics[width=0.48\textwidth]{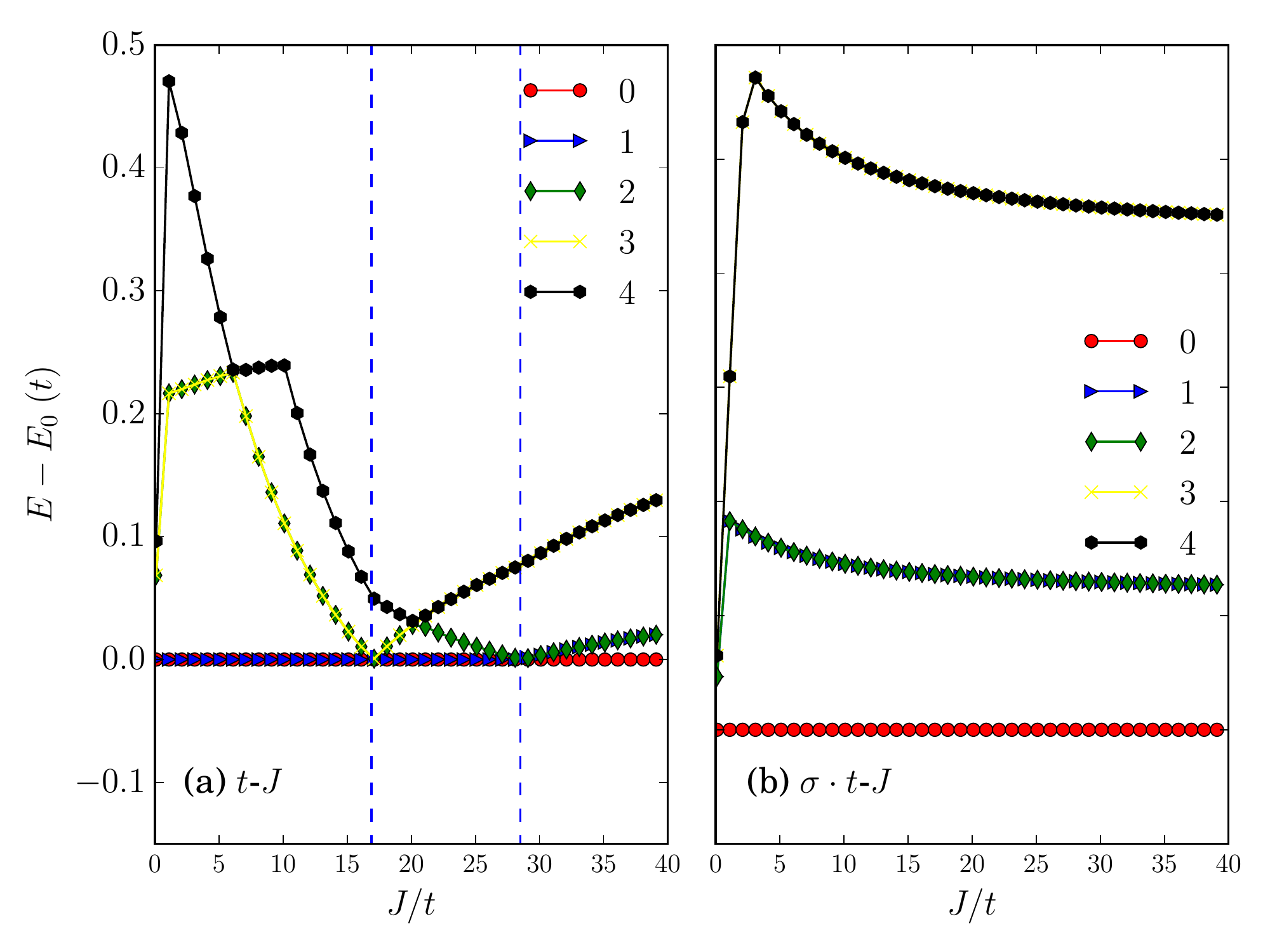}
\caption{The low-lying eigen-energies of the  $t$-$J$ model (a) and the $\sigma\cdot{t}$-$J$ model (b) (at $N=10$), respectively. The indices $0, 1, 2, \cdots$ label the energy levels. (a) The vertical dashed blue lines indicate two QCPs where the total momentum jumps; there is always a ground-state double degeneracy except for the trivial phase at $J>J_c$. (b) The ground state is always non-degenerate in the whole $J/t$ range. }
\label{fig:energy_len-10}
\end{figure}

Corresponding to the QCPs at the momentum jumps in Fig. \ref{fig:crystal_momenta} (a), the lowest two/three energy levels merge or cross as shown in Fig. \ref{fig:energy_len-10} (a) for the $t$-$J$ model. Consistent with the momentum characterization in Fig. \ref{fig:crystal_momenta} (a), the ground states are generically double-degenerate between QCPs except for the regime beyond the largest critical $J_c/t$ where the ground state reduces to a non-degenerate one.   

By comparison, Fig. \ref{fig:crystal_momenta} (b) indicates the corresponding phase diagram for the $\sigma \cdot t$-$J$ model, where the total momentum remains zero throughout the whole range of $J/t$. At the same time, the ground state always remains non-degenerate as shown in Fig. \ref{fig:energy_len-10} (b). Hence, the low-lying eigenstates of the $t$-$J$ and $\sigma \cdot t$-$J$ models look drastically different, except for the regime beyond the largest critical $J_c/t$ in the $t$-$J$ model. As pointed out in Introduction, the sole distinction between the $t$-$J$ and $\sigma \cdot t$-$J$ models lies in the sign structure. It thus confirms that the nontrivial QCPs and total momenta found in the $t$-$J$ model can all be attributed to the underlying phase-string sign structure. Note that in the regime of the trivial phase at $J>J_c$ in Fig. \ref{fig:crystal_momenta} (a), the ground state can be smoothly connected to the $t=0$ limit, where the hole becomes static and the phase-string sign structure of the $t$-$J$ model no longer functions, and only in this regime the two models predict the similar behavior or, in other words, they are adiabatically connected (see more discussion below).

\subsection{The momentum distribution of the electrons}

\begin{figure}[]
\centering
\includegraphics[width=0.48\textwidth]{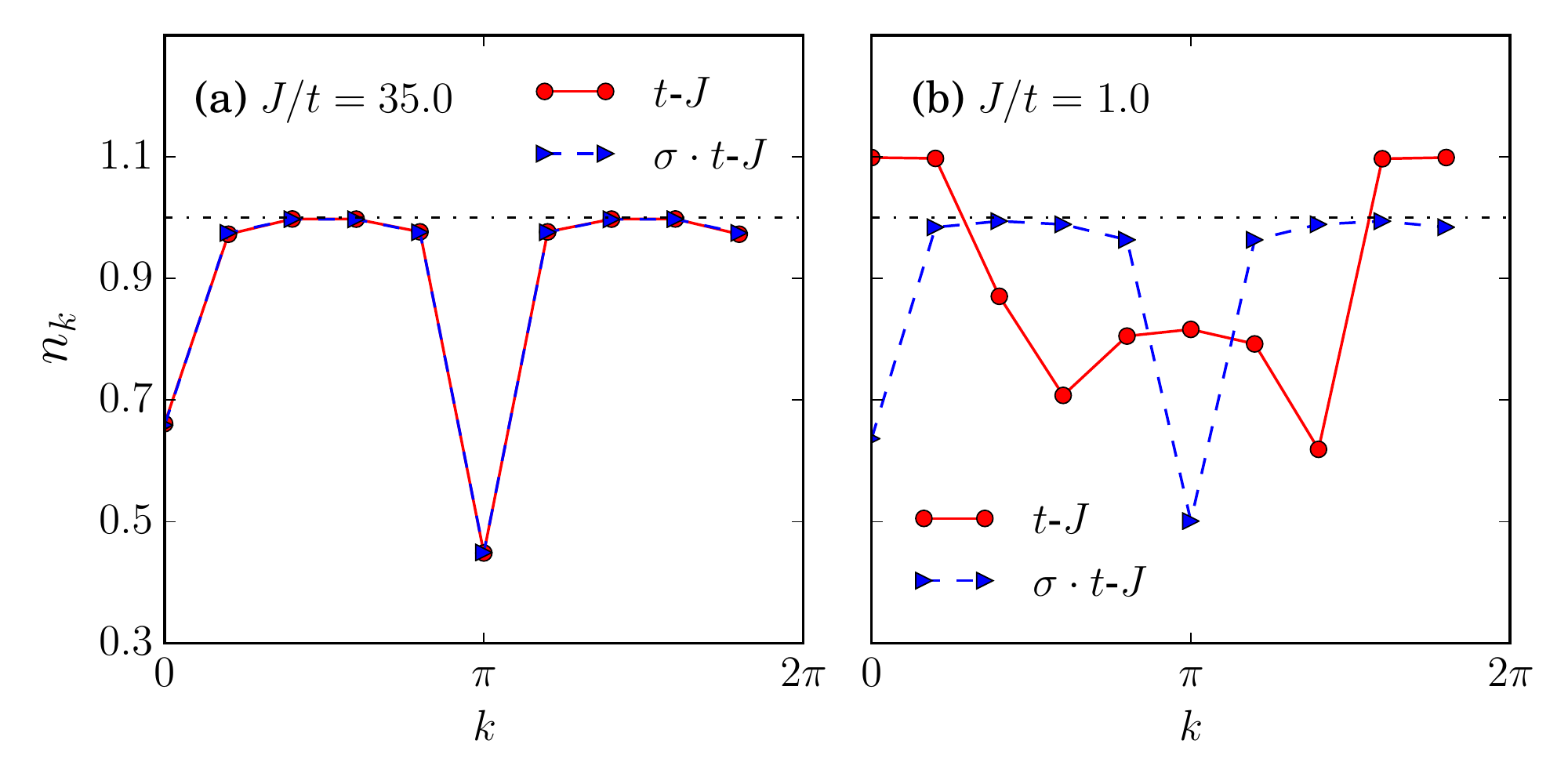}
\caption{Momentum distribution $n_{k}$ of the electrons at $N=10$ for the models of the $t$-$J$ (red circles) and the $\sigma\cdot{t}$-$J$ (blue triangles) in the one-hole ground state ($n_{k}=1$ at half-filling, the horizontal dot-dash  line). (a) $J/t=35.0>J_{c}/t$, $n_{k}$ peaks at $0$ and $\pi$ and is the same for both cases with the total momentum $K_0=0$; (b) $J/t=1.0$, a novel broad feature shows up in $n_{k}$ for the $t$-$J$ model at a given total momentum $K_0=2\pi/5$ while $n_{k}$ remains essentially unchanged for the $\sigma\cdot{t}$-$J$ model. }
\label{fig:nk_len-10}
\end{figure}

While the total momentum is conserved as marked in the phase diagram of Fig. \ref{fig:crystal_momenta}, the momentum distribution defined by
\begin{equation}
n_{k}=\sum_{\sigma}\langle\psi|c_{k\sigma}^{\dagger}c_{k\sigma}|\psi\rangle ~,
\label{}
\end{equation}
will provide further information on \emph{individual} electrons. In fact, at half-filling, each electron is localized at one lattice site such that $n_{k}=1$ [the horizontal dot-dash lines in Fig. \ref{fig:nk_len-10} where the $t$-$J$ and $\sigma\cdot{t}$-$J$ models are the same]. This is the most extreme case of strong correlation, i.e., the Mott insulator. Then, $1-n_{k}$ in the one-hole state effectively measures the change of the momentum distribution involving a doped hole in a loop of size $N$, which is essentially a many-body system involves $N-1$ electrons.

Upon one hole doping, the change of the momentum distribution is illustrated in Fig. \ref{fig:nk_len-10}. First, we note that $n_{k}$ is essentially the same in Fig. \ref{fig:nk_len-10} (a) for both $t$-$J$ and $\sigma\cdot{t}$-$J$ models at $J/t=35>J_c/t$ ($N=10$), where the total momentum is zero in the trivial phase according to Fig. \ref{fig:crystal_momenta}. Here $n_{k}$ peaks not only at $0$, but also at $\pi$, which may be explained by that a low-lying spin excitation carries away an antiferromagnetic wavevector $\pi$ as has been previously discussed \cite{zhu201651} in the large-$N$ limit.   

By contrast, $n_{k}$ of the $t$-$J$ model becomes qualitatively different from that of the $\sigma\cdot{t}$-$J$ model at $J/t=1$ in Fig. \ref{fig:nk_len-10} (b), corresponding to a total momentum $K_0=2\pi/5$ in Fig. \ref{fig:crystal_momenta} (a). Here, besides a peak at $K_0=2\pi/5$, another peak  (though smaller) also emerges at $-K_{0}$, and in addition, $n_{k}$ exhibits a ``continuum'' at all the allowed crystal momenta (i.e., $n2\pi/N$ with $n=0$, $\pm 1$, ...). Namely, given a conserved total momentum $K_0$, $n_{k}$ indicates that the momentum distribution of a single hole, $1-n_{k}$, is spread all over, which implies that the one-hole state is in a strongly correlated many-body ground state without a clear trace of the one-to-one correspondence as in a conventional quasiparticle picture. It is noted that even in the large-$N$ limit a similar ``continuum'' is still present as the hallmark of the Luttinger liquid behavior \cite{zhu201651}\cite{PhysRevB.41.2326}.


\subsection{The hole-spin correlations}

Next one may further investigate the many-body structure of the ground state in terms of hole-spin correlation functions defined below. Here the total spin quantum numbers of the one-hole ground state are $S^z=1/2$ and ${\bf S}^2=3/4$, with the longitudinal hole-spin correlator given by 
\begin{equation}
\begin{split}
C^{z}(h, j)&=\langle{n_{h}}S^{z}_{j}\rangle,
\end{split}
\label{}
\end{equation}
where $h$ and $j$ denote the sites of the hole and spin-$z$ component operators, respectively. 

\begin{figure}[h]
\centering
\includegraphics[width=0.4\textwidth]{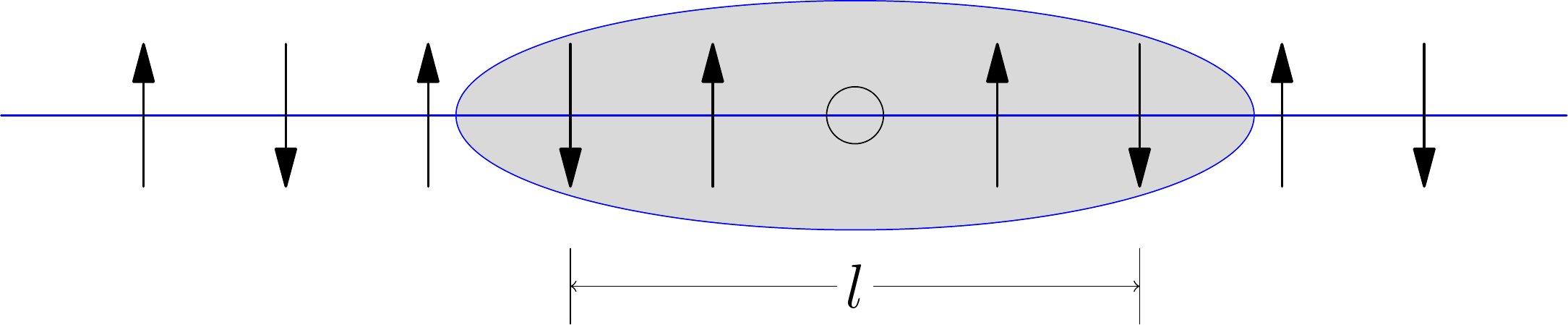}
\caption{Illustration of the region of size $l$ with the hole embedded at the center. (As a convention, if $l$ is an odd integer, the number of sites on the right side of the hole $h$ is larger than the left one by one.) The longitudinal and transverse parts of the total ${\bf S}^2$ measured within the regine $l$ are defined in Eqs. (\ref{quantum}).     
}
\label{fig:fixedhole}
\end{figure}
\begin{figure}
\centering
\includegraphics[width=0.48\textwidth]{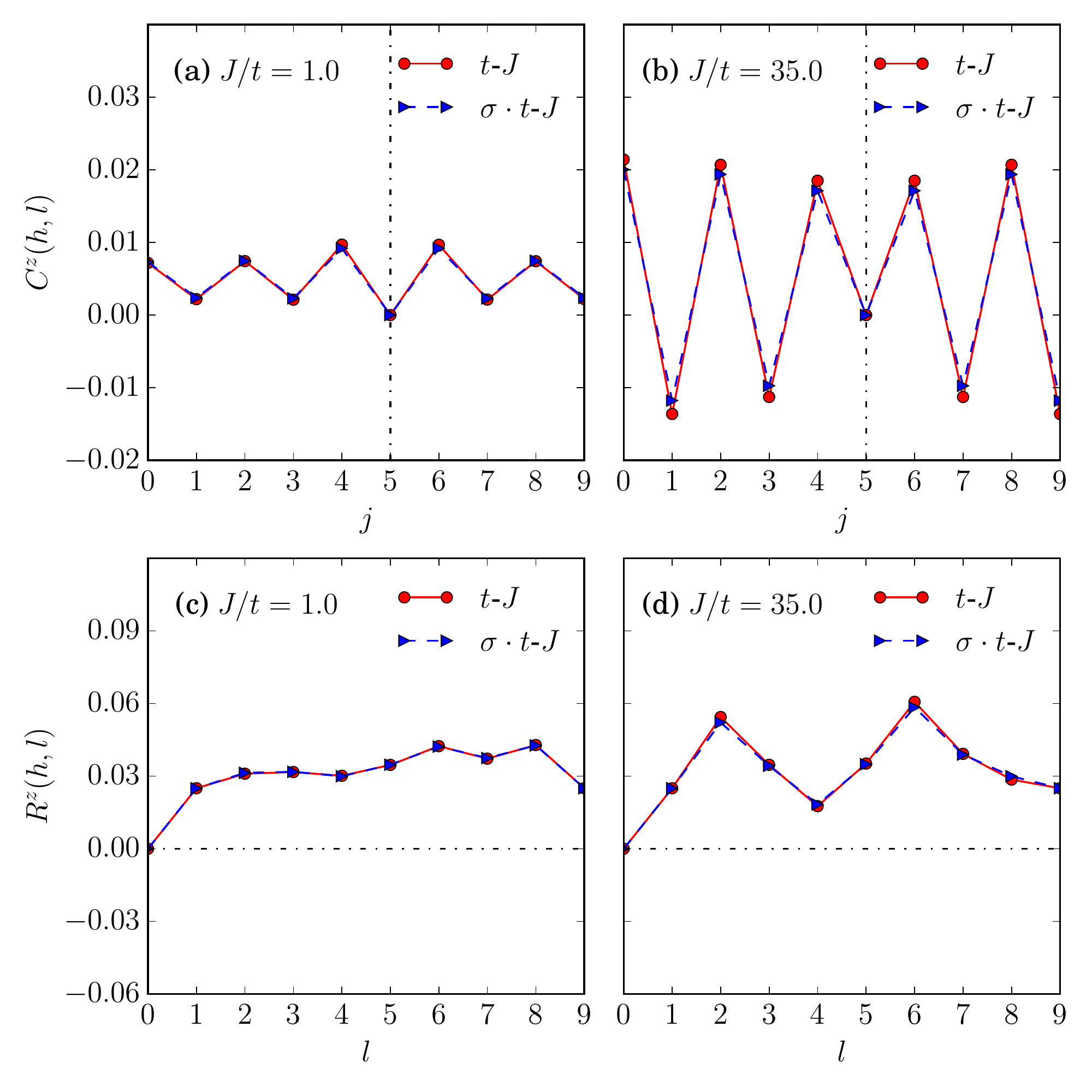}
\caption{There is no visible difference between the $t$-$J$ model and $\sigma\cdot{t}$-$J$ model with regard to the spin-$S_j^z$ distributions in the one-hole state with total $S^z=1/2$ ($N=10$). Here (a) and (b): the hole and spin-$S^z$ correlation $C^{z}(h, j)$ at $J/t=1.0$ and $J/t=35.0$, respectively. The vertical dashed lines mark the hole position; (c) and (d) : the computed $R^{z}(h, l)$.}
\label{fig:spinz_len-10}
\end{figure}

\begin{figure}
\centering
\includegraphics[width=0.48\textwidth]{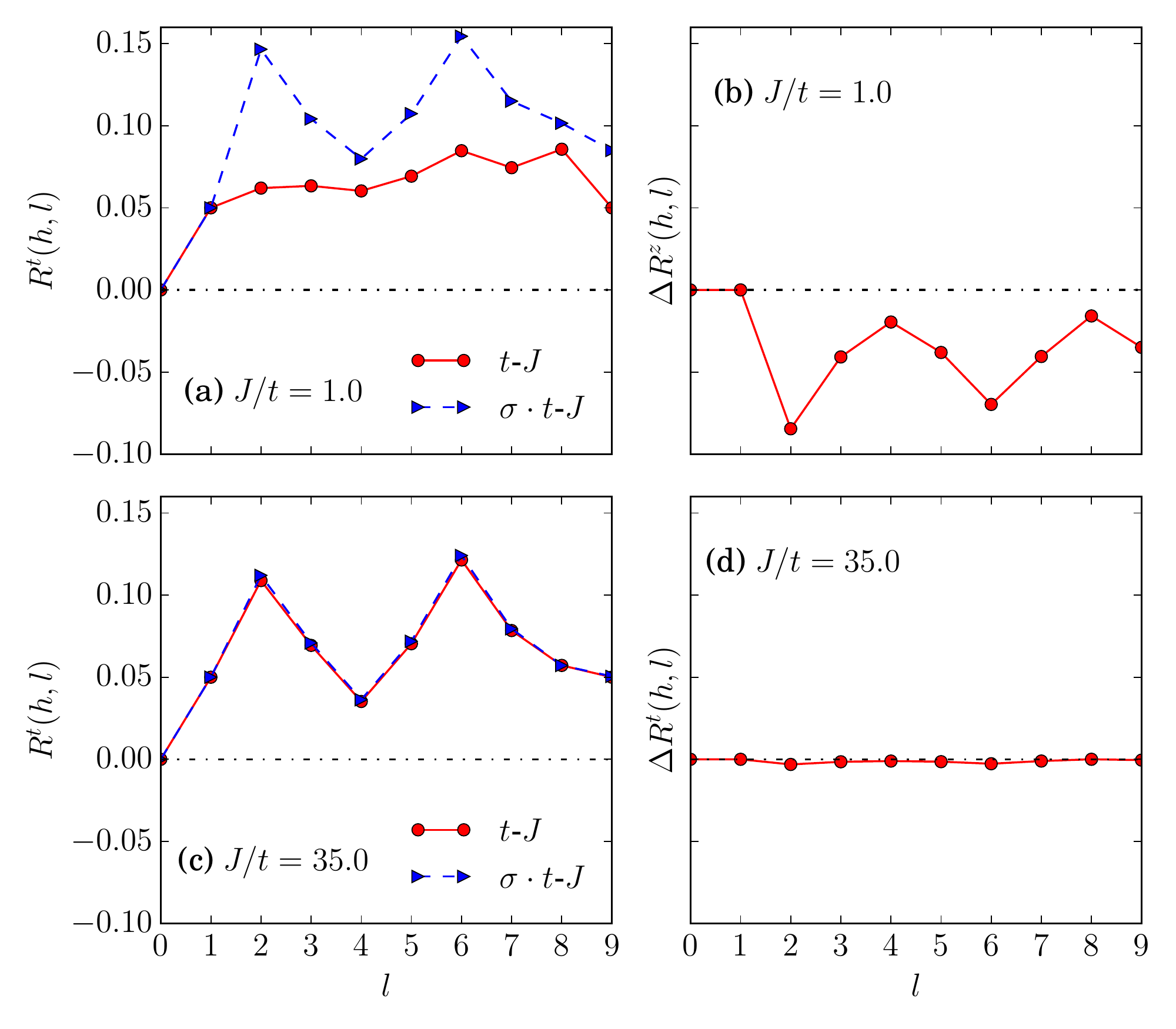}
\caption{The transverse component spin distribution $R^{t}(h, l)$ for the  $t$-$J$ and $\sigma\cdot{t}$-$J$ models and the net difference $\Delta R^{t}(h, l)$ between them, which indicate strong distinction [(a) and (b)] in the nontrivial phase $J/t=1.0$; but the distinction disappears [(c) and (d)] in the trivial phase $J/t=35.0>J_c/t$ ($N=10$).}
\label{fig:spint_len-10}
\end{figure}

In order to measure the distribution of the quantum fluctuation of ${\bf S}^2=3/4$ around the hole, we further introduce the following correlators 
\begin{equation}
\begin{split}
R^{z}(h, l)&=\sum_{(i, j)\in{l}}\langle{S}^{z}_{i}S^{z}_{j}\rangle \\
R^{t}(h, l)&=\sum_{(i, j)\in{l}}\frac{1}{2}\langle({S}^{+}_{i}S^{-}_{j}+h.c.)\rangle \\
\Delta{R}^{z, t}(h, l)&={R}^{z, t}_{t\text{-}J}(h, l)-R^{z, t}_{\sigma\cdot{t}\text{-}J}(h, l)
\end{split}
\label{quantum}
\end{equation}
where $R^{z,t}(h, l)$ describe the longitudinal and transverse fractions of the total ${\bf S}^2=3/4$ within the region $l$ with the hole embedded at the center $h$ as illustrated in Fig. \ref{fig:fixedhole}. 

The longitudinal spin distributions around the hole $h$ fixed at the center of the sample are shown in Fig. \ref{fig:spinz_len-10} for $C^{z}(h, j)$ and $R^{z}(h, l)$, respectively. Even though these correlators differ at different $J/t$'s, a prominent feature is that there is no obvious difference between the $t$-$J$ and $\sigma\cdot{t}$-$J$ models for the longitudinal spins correlated with the hole.

However, a sharp difference between the two models shows up in the transverse component $R^{t}(h, l)$ in Fig. \ref{fig:spint_len-10} (a) or $\Delta{R}^{t}(h, l)$ in Fig. \ref{fig:spint_len-10} (b) at $J/t=1$. By contrast, the distinction disappears at a larger $J/t$, say, $35>J_c/t$ in Figs. \ref{fig:spint_len-10} (c) and (d), which is in the trivial non-degenerate regime of Fig. \ref{fig:crystal_momenta}.  

Since the $t$-$J$ and $\sigma\cdot{t}$-$J$ models only differ by a phase-string sign structure, the above hole-spin correlators suggest that such a distinction is intrinsically and solely related to the \emph{transverse component} of the quantum fluctuation ${\bf S}^2=3/4$ around the hole. In other words, the nontrivial QCPs, the double degeneracy of the ground state, the total momenta, as well as the anomalous momentum distribution of the hole in the $t$-$J$ model can all be traced back to the transverse quantum fluctuation in ${\bf S}^2=3/4$, which distributes differently from the $S^z=1/2$ component. Here, the phase-string sign structure is a precise mathematical description of such a novel quantum fluctuation effect induced by doping. By contrast, at a sufficiently large $J/t>J_c/t$, we have seen that $R^{z, t}(h, l)$ behave similarly, implying that the phase string effect gets ``screened'', where the longitudinal $S^z=1/2$ and ${\bf S}^2=3/4$ can be combined to be described by an integral $S=1/2$, which is loosely detached from the hole in the 1D chain due to a long-range spin-spin correlation.    

\section{Spin-charge mutual entanglement}

In the previous section, the one-hole state of the $t$-$J$ loop has been precisely studied by ED, which has revealed a series of novel many-body properties manifested in the phase diagram, described by nontrivial total (many-body) momentum, QCPs, ground state degeneracy, the momentum distribution continuum of the single hole, and the distinct spin-spin correlation induced by the hole.  However, the above characterizations of these properties look quite detailed, involving all different kinds of conventional correlation functions. 
It is natural to ask if one may design a unified description to capture the essential physics in such a strongly correlated system.

It is seen above that a doped hole does not simply dissociate into a holon and a spin-1/2 spinon in the $t$-$J$ loop. Namely, the total spin $S^z=1/2$ and the transverse quantum fluctuation in ${\bf S}^2=3/4$ may be distributed quite differently in space, indicating that the many-body spins in the background must be also ``mingled'' in. The \emph{additional} transverse spin-spin correlation induced by the motion of the hole results in a novel phase-string sign structure in the $t$-$J$ model, and it is the quantum interference of the latter in the closed loop leads to the aforementioned strongly correlated properties. Therefore, how to capture such a nonlocal ``entanglement'' between the doped charge and the rest of spin degrees of freedom is the key here. 




As a first step, we may expand the ground state in a \emph{direct product representation} of the two components which we are interested in. As discussed in detail in Appendix \ref{sec:appendix_entanglement}, these two components must be ordered and disentangled in such a way that they cannot retrieve each others' quantum information from the basis state alone. The natural choice of the basis state can be then a direct product $|h\rangle\otimes|\{s\}\rangle$, where the spinless hole (holon) state $|h\rangle$ is at any site $h$ of the original $t$-$J$ loop, while $|\{s\}\rangle$ an arbitrary spin configuration of $N-1$ sites as a subsystem with the hole site excluded. It is important to note that in $|\{s\}\rangle$ the hole site $h$ is excluded like in the so-called ``squeezed spin chain'' \cite{PhysRevB.41.2326}\cite{PhysRevB.70.075109}. 



 
Then the ground state wave function can be written as
 \begin{equation}
 |\psi\rangle=\sum_{h, \{s\}}c_{hs}|h\rangle\otimes|\{s\}\rangle ,
 \label{direct product}
 \end{equation}
where the nontrivial mutual correlation between the hole and its surrounding spins or the hole-spin mutual entanglement, will be uniquely encoded in the expansion coefficient $c$ in Eq. \ref{direct product}. 
 
 

By tracing out the spin configurations, one gets an $N\times{N}$ reduced density matrix $\rho_{h}$ for the holon (charge):
\begin{equation}
(\rho_{h})_{ab}=\sum_{\{s\}}c_{as}c_{bs}^{*}.
\label{}
\end{equation}
Correspondingly the charge-spin MEE can be defined as a von Neumann entropy as follows
\begin{equation}
    S_h = -\text{tr}(\rho_{h}\ln\rho_{h})=\sum_{k=0}^{r-1}|\lambda_{k}|^{2}\ln|\lambda_{k}|^{2},
    \label{eq:entanglement_entropy}
\end{equation}
where $r$ is the rank of the reduced density matrix and $\lambda_{k}$ represents the complex amplitude of the wave function in the transformed basis. For more details, please refer to Appendix \ref{sec:appendix_entanglement}.

\begin{figure}[]
\centering
\includegraphics[scale=0.4]{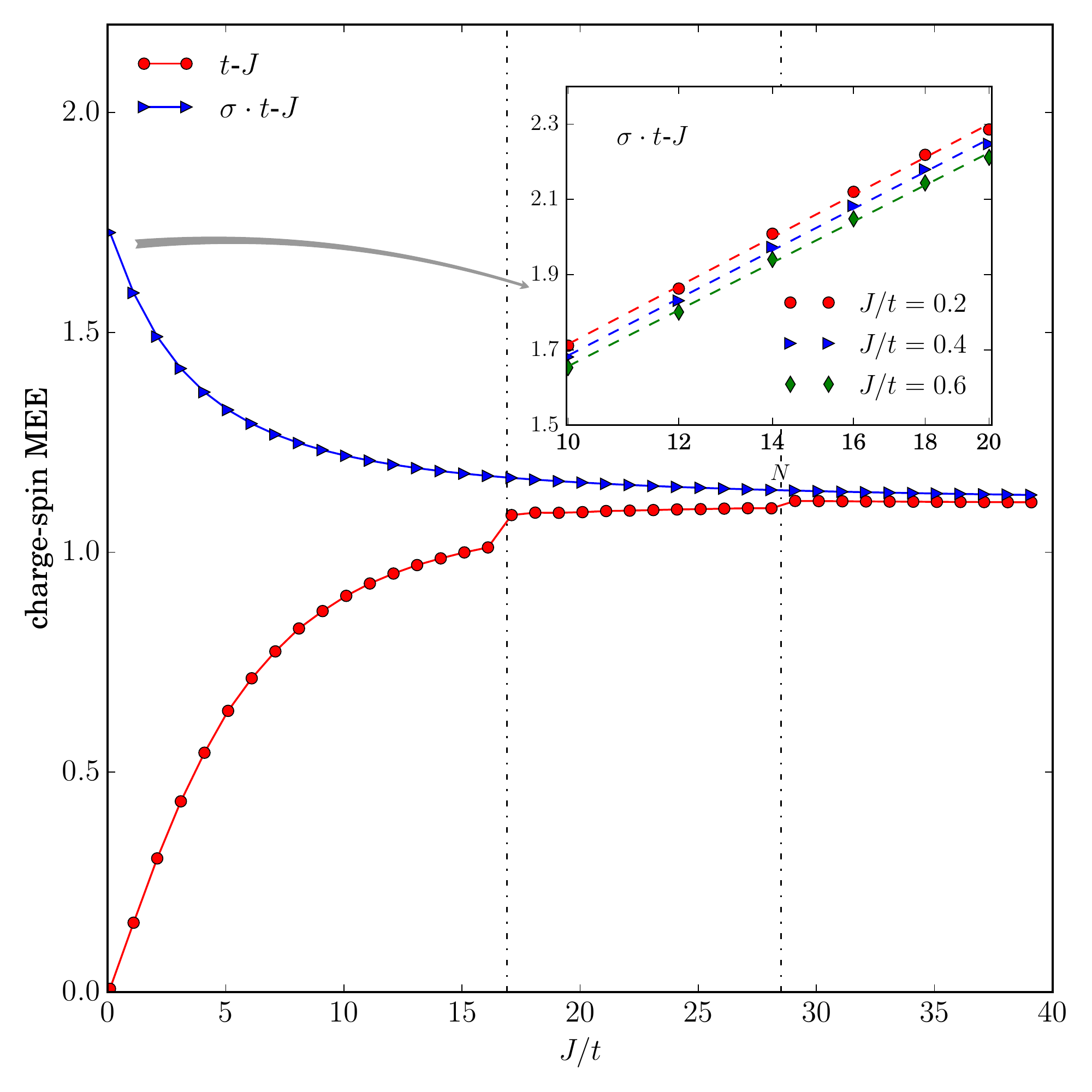}
\caption{Spin-charge mutual entanglement entropy (MEE) for both the $t$-$J$ and $\sigma\cdot{t}$-$J$ loops at $N=10$. Inset: the finite-size scaling of the MEE for the $\sigma\cdot{t}$-$J$ loop well follows a $\log(N)$ behavior in the small $J/t$ limit. But the MEE of the $t$-$J$ loop shows an opposite trend in the same limit, which vanishes in consistency with the spin-charge separation.}
\label{fig:entanglement_entropy_len-10}
\end{figure}

The numerical results of the spin-charge MEE are shown in Fig. \ref{fig:entanglement_entropy_len-10}. Note that for the degenerate ground states of the $t$-$J$ model, we have chosen the positive total momentum $K_0$ without loss of generality. As Fig. \ref{fig:entanglement_entropy_len-10} shows, the QCPs in the phase diagram of Fig. \ref{fig:crystal_momenta} are clearly exhibited by distinct jumps in the MEE of the $t$-$J$ model. With the increase of $J/t$, eventually the difference between the $t$-$J$ model and $\sigma\cdot{t}$-$J$ models disappears in the non-degenerate regime (with $J/t>(J/t)c= 28.5$ at $N=10$). 

In the opposite limit of small $J/t$, the distinction between the $t$-$J$ and $\sigma\cdot{t}$-$J$ models gets progressively enlarged. Actually the spin-charge MEE approaches opposite limits for the two cases. In the $t$-$J$ loop case, it vanishes as $J/t\rightarrow 0$, in consistency with the true spin-charge separation. It has been well established \cite{PhysRevB.55.3894}\cite{PhysRevB.70.075109} that when the hopping is much faster than the superexchange process, the holon is effectively decoupled from the background ``squeezed spin chain'' described by a Heisenberg model of $N-1$ sites. 

By contrast, the spin-charge MEE for the $\sigma\cdot{t}$-$J$ model increases and reaches a maximum at $J/t\rightarrow 0$. At a given small $J/t$, the finite-size scaling of the MEE is illustrated in the insert of Fig. \ref{fig:entanglement_entropy_len-10}. It shows a nice $\log(N)$ scaling behavior similar to that of the conventional EE predicted by conformal field theory\cite{fradkin2013QFT} for a pure 1D Heisenberg spin chain.

Therefore, the opposite trends of the MEE at small $J/t $ demonstrate that the presence/absence of the phase-string sign structure fundamentally influence the underlying hole-spin correlation from the spin-charge separation in the $t$-$J$ model to the most strong hole-spin entanglement in $\sigma\cdot{t}$-$J$ model. In the latter case, the hole behaves as if it is still a spin in the original undoped Heisenberg model, maintaining similar long-range RVB correlations with the rest of spins.       





\begin{figure}[]
\centering
\includegraphics[scale=0.4]{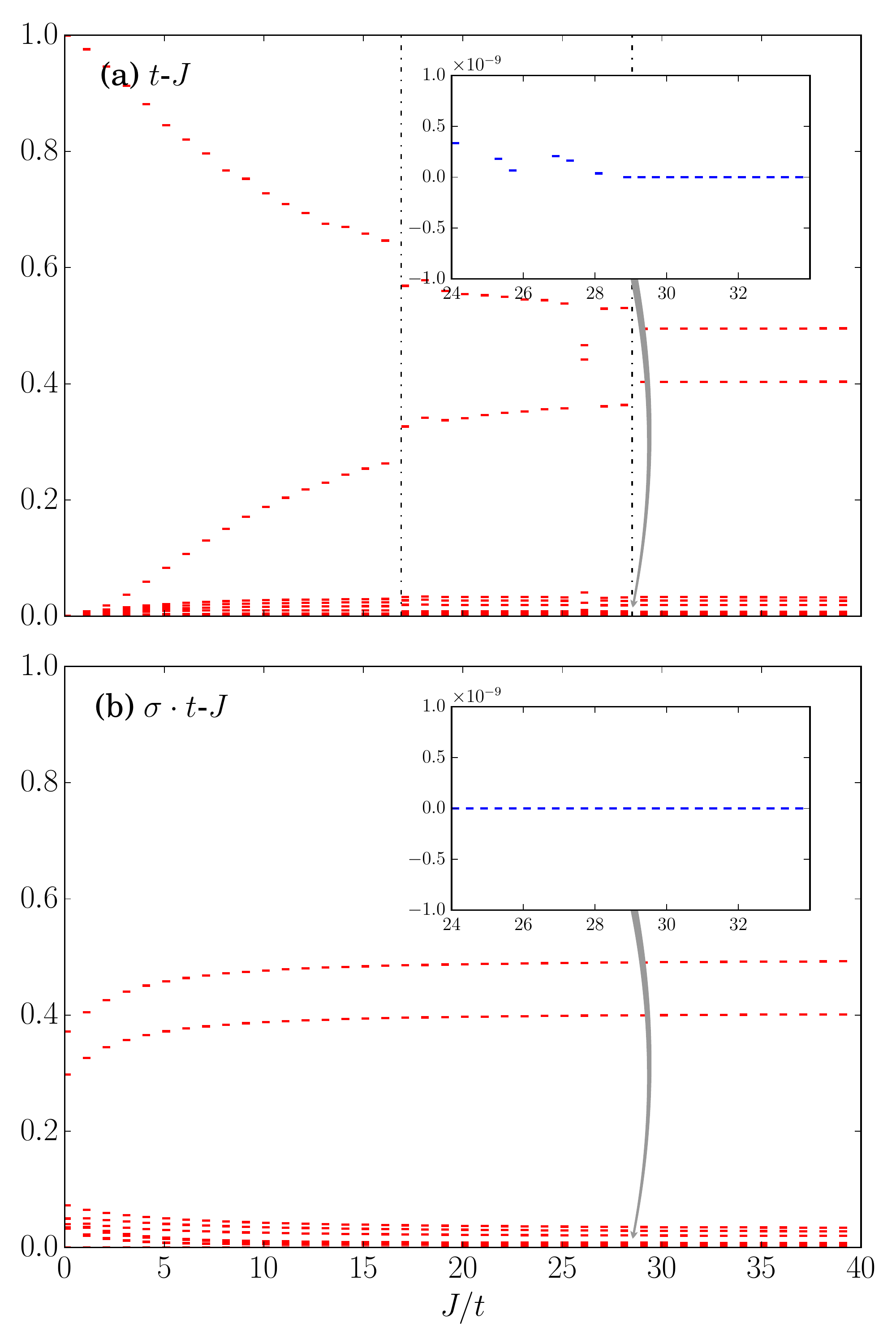}
\caption{The spin-charge mutual entanglement spectrum (MES) at $N=10$:  (a) the $t$-$J$ model and (b) the $\sigma\cdot{t}$-$J$ model. }
\label{fig:entanglement_spectrum_len-10}
\end{figure}


In Fig. \ref{fig:entanglement_spectrum_len-10}, the entanglement spectrum MES defined as $\{|\lambda_{k}|^{2}\}$ in Eq. \ref{eq:entanglement_entropy} is shown for the $t$-$J$ (a) and $\sigma\cdot{t}$-$J$ (b) models, respectively. Like the MEE, the MES also shows distinct signatures at QCPs in Fig. \ref{fig:entanglement_spectrum_len-10} (a). Note that the lowest eigenvalue remains at zero in both the $t$-$J$ model and $\sigma\cdot{t}$-$J$ models in the large $J/t$ trivial region. However, once in the nontrivial regimes at smaller $J/t$ (below the largest QCP $(J/t)_c$), the lowest entanglement eigenvalue becomes always gapped as shown in the insert of Fig. \ref{fig:entanglement_spectrum_len-10} (a) in the $t$-$J$ model. In particular, the gap seems randomly fluctuates versus $J/t$. As discussed in the previous section, in the same regime, the hole becomes incoherent as a ``continuum'' of momentum is involved. 

\section{Conclusion and discussion}

In this work, we have used an exact numerical method to study one of the simplest systems of doped Mott insulators, i.e., a single hole in a finite-size spin loop described by the $t$-$J$ Hamiltonian. One may imagine that the ground state would be intrinsically complicated with the single hole strongly entangled with the background spins. To effectively characterize such a strongly correlated problem, we have introduced a mutual entanglement description, i.e., the charge-spin mutual entanglement entropy. Based on this new quantity calculated by ED, a series of QCPs of the ground state along the axis of $J/t$ have been revealed. In particular, the trivial case is at $t=0$ where the hole is static and the mutual entanglement reduces to that of the half-filling spin chain partitioned between a spin at the hole site and the rest of spins. Then, with increasing $t/J$, the mutual entanglement is monotonically reduced, with the sharp jumps at the QCPs,  and eventually vanishes at $J/t\rightarrow 0$. The latter is nothing but the limit of spin-charge separation. Further information is provided by the mutual entanglement spectrum, which in the trivial phase at large $J/t$ can be smoothly connected to the $t=0$ limit, but the lowest eigenvalue becomes finite and seems randomly fluctuating with $J/t$ in the so-called nontrivial regimes bounded by the QCPs at smaller $J/t$.    

However, once the so-called phase-string sign structure is turned off in the model, the above anomalous behaviors all disappear. Namely, all the QCPs no longer exist, with the mutual entanglement entropy becomes a smooth function, which increases with reducing $J/t$ and eventually reaches a \emph{maximum}, as opposing to \emph{vanishing} in the presence of the phase-string sign structure, at $J/t\rightarrow 0$. Thus, in contrast to the spin-charge separation, without the phase string, the hole-spin entanglement is maximized as if there is still a spin at the hole site (that hops quickly from site to site), which well retains the same correlations with the rest of spins as at half-filling where all spins form resonating-valence-bond type of singlet pairing. 

Indeed, the phase string counts how the motion of the hole scrambles the Marshall signs\cite{PhysRevLett.77.5102}\cite{marshall1955antiferromagnetism} that ensures the transverse part of the singlet pairing of spins at half-filling. The ED results have explicitly shown that the phase-string sign structure is directly associated with the transverse component of ${\bf S}^2=3/4$. Namely, in the one-hole ground state of a bipartite lattice, the total spin is $S=1/2$. However, it does not mean that one can construct the ground state in terms of a spinless holon and a spinon in a naive spin-charge separation picture. Rather, the background spins will mingle with the spinon, in the form of phase string or many-body phase shift, such that the longitudinal $S^z=1/2$ and the transverse part of ${\bf S}^2=3/4$ can no longer be described by a single S=1/2 spinon in general. As the consequence, the conserved many-body total momentum is shared by the hole and the spins as well. In particular, the transverse spin component strongly smears the momentum distribution to render the charge incoherent or translation symmetry breaking since it is not a rigid part of the charge. As a function of $J/t$, a sequence of the total momenta can be thus generated in the ground state with level crossing at QCPs. Again, once the phase string is turned off, the ground state always remains trivial with the doped hole properly described in terms of a composite pair of holon and spinon of $S=1/2$.   

It is important to note that the above spin-charge mutual entanglement description is not restricted in a 1D loop. The algorithm can be generalized to the 2D case shown in Fig. \ref{fig:illustration}, in which the 1D loop only constitutes just some special closed path in the whole summation of, say, the partition function defined in Eq. (\ref{Ztj}). But even in the present extremely simplified case, the lesson is clear. That is, in the doped Mott insulator, the many-body quantum mechanism may have no classical limit and the quantum interference of the singular sign (phase) effect from \emph{all} paths has to be fully considered. As such, the mutual entanglement between the charge and spin degrees of freedom may become an essential characterization of the underlying physics.    



Finally, we point out that the present results may have some profound relation with the idea of quantum disentangled liquids (QDL) \cite{Grover:P10010}\cite{1611.02075}. A QDL is conjectured to be a new kind of quantum fluid composed of two components of light and heavy particles. The essence of QDL is that the light particles may not be able to thermalize because their highly excited states are in the many-body localization (MBL)\cite{annurevMBL} states in the presence of the heavy particles which are thermalized \cite{Grover:P10010}\cite{1611.02075}. Due to the disentanglement between the two components, eigenstate thermalization hypothesis (ETH) may be violated there. Even though the present work deals with small size systems with one hole near the ground state, the mutual entanglement entropy clearly indicates a vanishing trend in small $J/t$ limit, where the charge is like a light particle while the spins are heavy particles. An area law (constant) behavior is also found in the finite-size scaling of $N$ in this regime. In other words, our results may have shown the precursor of a prototypical QDL composed of incoherent light charges and heavy spins in the $t$-$J$ model, where the ETH may fail at finite doping and finite energy density states. Recently, the signature of a possible MBL has been also found in the study of an extremely large $U$ model of two-leg ladder by the density matrix renormalization group method. Interestingly, when the phase-string sign structure is turned off, all the signatures of the MBL disappear completely\cite{ScientificReports.6.35208}.

\begin{acknowledgements}
 We acknowledge stimulating discussions with X. Chen, M.P.A. Fisher, Y.-M. Lu, X.-L. Qi, Y. Qi, Q.-R. Wang, P. Ye, Y.-Z. You,  J. Zaanen, and Z. Zhu. This work is supported by NSFC and MOST of China. 
\end{acknowledgements}

\appendix

\section{The precise sign structure of the 1D $t$-$J$ loop with one hole }
\label{sec:intro}

The 1D $t$-$J$ Hamiltonian is defined in the Hilbert space with projecting out double-occupancy, which reads $H = H_{t}+H_{J}$ where
\begin{equation}\label{tj}
\begin{split}
H_{t} &= -t\sum_{\langle{ij}\rangle, \sigma}(c_{i\sigma}^{\dagger}c_{j\sigma}+h.c.), \\
H_{J} &= J\sum_{\langle{ij}\rangle}\left(\mathbf{S}_{i}\cdot\mathbf{S}_{j}-\frac{1}{4}n_{i}n_{j}\right).
\end{split}  
\end{equation}
A finite-size loop of $N$ sites with $N-1$ spin-$1/2$ electrons created by the operator $c^{\dagger}_{\sigma}$ will be considered. The summations in Eq. (\ref{tj}) run over all the nearest-neighbors $\langle{i, j}\rangle$. In the undoped case (half-filling), the model reduces to a Heisenberg spin chain, with the ground state satisfying the Marshall sign rule\cite{marshall1955antiferromagnetism}. Define the Marshall-sign basis $|\{s\}\rangle = (-)^{N_{A}^{\downarrow}}c_{0\sigma_{0}}^{\dagger}\cdots{c}_{N-1\sigma_{N-1}}^{\dagger}|0\rangle$ in which $N_{A}^{\downarrow}$ denotes the total number of down spins at sub-lattice $A$ in an even-$N$ loop. In this basis  $\langle\{s{'}\}|H_{J}|\{s\}\rangle\leq{0}$ and Perron-Frobenius theorem implies that the ground state at half-filling can be written as $|\psi_{0}\rangle=\sum_{\{s\}}a_{s}|\{s\}\rangle, a_{s}\geq{0}$. Generalized to doped case, $|\{s\}, h\rangle=(-)^{h}c_{h\sigma}|\{s\}\rangle$, the dilemma arises immediately because of $\langle\{s\}, h|H_{t}|\{s{'}\}, h{'}\rangle=-t\sigma_{h}$. That is, the hopping process will violate the Marshall sign rule to cause an irreparable many-body phase shift known as the phase string \cite{PhysRevB.55.3894}.

In order to remove the sign frustration caused by the hopping, the so-called $\sigma\cdot{t}$-$J$ model may be introduced in which without changing $H_J$, the hopping term $H_t$ is modified to
\begin{equation}\label{stj}
H_{\sigma\cdot{t}} = -t\sum_{\langle{i, j}\rangle, \sigma}\sigma(c_{i\sigma}^{\dagger}c_{j\sigma}+h.c.),
\end{equation}
where an extra \text{spin-dependent} sign $\sigma=\pm$ is introduced which can precisely erase the non-local phase string effect. Then one can obtain a sign-free basis for the $\sigma\cdot{t}$-$J$ model as\cite{zhu201651}:
\begin{equation}
    |\psi\rangle_{\sigma\cdot{t}\text{-}J}=\sum_{h, \{s\}}a_{h, \{s\}}(-\sigma)^{h}c_{h\sigma}|\{s\}\rangle,
\end{equation}
where $\sigma$ denotes the spin of the electron removed from the system and $a_{h,\{s\}}$ is always \emph{positive} such that the sign structure of  the $\sigma\cdot{t}$-$J$ model is trivial. According to the Perron-Frobenius theorem, there is no ground state degeneracy in this case. 

\begin{figure}
\centering
\includegraphics[width=0.45\textwidth]{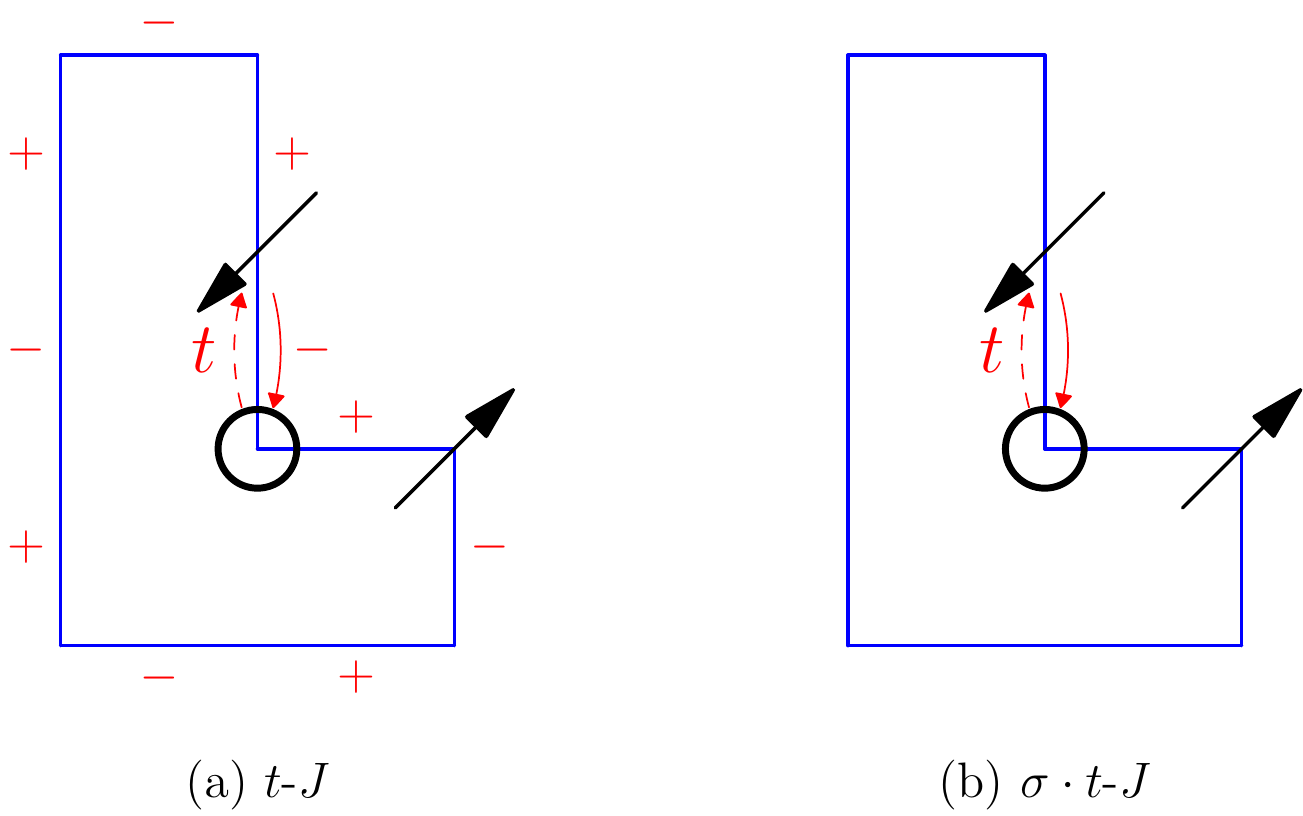}
\caption{(a) Phase string $(+)\times(-)\times(-)\cdots$ picked up by the hole in the $t$-$J$ model along a specific path plays the crucial role in strongly correlated doped Mott physics. (b) In the $\sigma\cdot{t}$-$J$ model, the phase string frustration is precisely removed.}
\label{fig:phasestring}
\end{figure}

Generically it has been proved\cite{PhysRevB.77.155102}\cite{PhysRevLett.77.5102} rigorously that in a single hole doped $t$-$J$ model, the propagating hole along any specific path $c$ will pick up the spin-dependent signs as $(+)\times(-)\times(-)\cdots$. The signs $\pm$ denote the quantum memory of the microscopic hopping process by the hole, which records whether it exchanges with an $\uparrow$ or $\downarrow$-spin. Phase string effect is illustrated as in Fig. \ref{fig:phasestring}(a). The exact sign structure for a single hole is $\tau_{c}=(-1)^{N_{h}^{\downarrow}[c]}$, where $N_{h}^{\downarrow}[c]$ counts the total number of exchanges between the hole and down spins along the closed path $c$,
\begin{equation}
Z_{t\text{-}J}=\sum_{c}\tau_{c}\mathcal{W}[c],
\label{}
\end{equation}
in which $\mathcal{W}[c]\geq{0}$. By comparison, the partition function of the corresponding $\sigma\cdot{t}$-$J$ model is given by $Z_{\sigma\cdot{t}\text{-}J}=\sum_{c}\mathcal{W}[c]$, where $\tau_{c}$ is precisely removed with the same positive weight $\mathcal{W}[c]$. In other words, the $t$-$J$ model and $\sigma\cdot{t}$-$J$ model solely differ by the phase-string sign structure.  \\

\section{Some details}
\subsection{Hilbert space construction}
To construct the Hilbert space of $t$-$J$ model means that we should construct the basis of this space. Naively we refer to all kinds of possible configurations. In the first place, you should take care that this is a fermionic problem while in practice we use bosonic representation of a configuration in the computer which means that an order of the fermionic operators should be assumed. For instance,
\begin{equation}
|s_{0}^{z}, ..., s_{h-1}^{z}, o, s_{h+1}^{z}, ..., s_{N-1}^{z}\rangle\equiv{c}_{0}^{\dagger}...{c}_{h-1}^{\dagger}{c}_{h+1}^{\dagger}...{c}_{N-1}^{\dagger}|0\rangle,
\end{equation}
where a single hole is left on the site $h$ and the spin indices of the creation operators are ignored. In this sense you can check straightforwardly that the hopping term in the Hamiltonian acting on these kind of states has two distinct cases:
\begin{itemize}
\item Hopping within the chain, there is \textit{no} sign arised.
\item Hopping on the boundary for periodic boundary condition may give rise to another fermionic sign.
\end{itemize}
\begin{widetext}
\begin{equation}
(c_{N-1}^{\dagger}c_{0}+h.c.)c_{0}^{\dagger}...c_{N-2}^{\dagger}|0\rangle=c_{N-1}^{\dagger}c_{0}c_{0}^{\dagger}...c_{N-2}^{\dagger}|0\rangle=(-)^{N-2}c_{1}^{\dagger}...c_{N-1}^{\dagger}|0\rangle.
\end{equation}
\end{widetext}

Furthermore,
\begin{widetext}
\begin{equation}
|\varphi(h, s)\rangle\equiv(-\sigma)^{h}c_{h\sigma}|s\rangle=(-)^{n_{A}^{\downarrow}}\cdot(\sigma)^{h}c_{0\sigma_{0}}^{\dagger}\cdots{c}_{h\sigma}c_{h\sigma}^{\dagger}\cdots{c}_{N-1\sigma_{N-1}}^{\dagger}|0\rangle,
\end{equation}
\end{widetext}
in which $c_{0\sigma_{0}}^{\dagger}\cdots{c}_{h\sigma}c_{h\sigma}^{\dagger}\cdots{c}_{N-1\sigma_{N-1}}^{\dagger}|0\rangle$ is nothing but the convention of the basis in our practical computational program. Then we can compute the elements of correlation matrix which is useful for deriving the momentum distribution of the hole $n_{k\alpha} = \langle\psi|c_{k\alpha}^{\dagger}c_{k\alpha}|\psi\rangle_{\sigma\cdot{t}\text{-}J}$.
\begin{equation}
\langle\psi|c_{i\alpha}^{\dagger}c_{j\alpha}|\psi\rangle_{\sigma\cdot{t}\text{-}J} = \sum_{h, s; h', s'}a_{h,s}^{*}a_{h', s'}, \quad \alpha = \sigma, \bar{\sigma}.
\end{equation}
Here $\sigma$ denotes the spin of the electron removed from the half-filled background. It is 

\subsection{Translational operation}
\label{sec:translation_operator}
Essentially we define $Tc_{i}^{\dagger}T^{-1}\equiv{c}_{i+1}^{\dagger}$ and it commutes with the Hamiltonian in periodic boundary condition. It is obvious $T$ commutes with $H_{J}$ because $H_{J}$ denotes purely bosonic operations and they only acts on spins of electrons along the cyclic chain while $T$ only moves electrons to other sites. $T$ also commutes with  $H_{t}$. To see this,
 \begin{equation}
 \begin{split}
 TH_{t}T^{-1} &= -t\sum_{i=0}^{N-1}(Tc_{i}^{\dagger}T^{-1}Tc_{i+1}T^{-1}+Th.c.T^{-1}) \\
 &= -t\sum_{i=0}^{N-1}(c_{i+1}^{\dagger}c_{i+2}+h.c.) \\
 &= -t\sum_{i=0}^{N-1}(c_{i}^{\dagger}c_{i+1}+h.c.)
 \end{split}
 \end{equation}
 because of $i$'s modular definition along the cyclic chain. In this sense, $T$ may also give rise to some sign when it operates on a state. For example,
\begin{widetext}
\begin{equation}
Tc_{1}^{\dagger}...c_{N-1}^{\dagger}|0\rangle = Tc_{1}^{\dagger}T^{-1}Tc_{2}^{\dagger}...Tc_{N-1}^{\dagger}T^{-1}T|0\rangle = (-)^{N-2}c_{0}^{\dagger}...c_{N-1}^{\dagger}|0\rangle,
\end{equation} 
\end{widetext}
where we have assumed that the vacuum is invariant under the translation. You can also find that the only case in which there is no sign derived is that the hole is located on the site $N-1$. 

\section{Reduced density matrix and entanglement entropy}
\label{sec:appendix_entanglement}
\subsection{Composite system and reduced density matrix}
Consider a composite quantum system $\Sigma=A\cup{B}$ and suppose there are sets of complete basis $\{|i\rangle_{A}\}$ and $\{|j\rangle_{B}\}$ to span Hilbert spaces $\mathcal{H}_{A}$ and $\mathcal{H}_{B}$ with dimension $d_{A}$ and $d_{B}$, respectively. For the composite system $\Sigma$ the \emph{direct product} (Cartesian product) states $|i\rangle_{A}\times|j\rangle_{B}\equiv|i\rangle_{A}|j\rangle_{B}$ form a complete basis for the tensor product space $\mathcal{H}_\Sigma=\mathcal{H}_{A}\otimes\mathcal{H}_{B}$. This is called \emph{direct product representation} while a pure state in $\mathcal{H}_{\Sigma}$ generally cannot be written as a single product state but the superposition of them since there may exist quantum entanglement:
\begin{equation}
|\psi\rangle_{\Sigma}=\sum\limits_{i,j}c_{ij}|i\rangle_{A}|j\rangle_{B}.
\label{}
\end{equation}
Note that each Cartesian term $|i\rangle_{A}|j\rangle_{B}$ sometimes is written as $|i\rangle_{A}\otimes|j\rangle_{B}$\footnote{Strictly speaking, this very symbol $\otimes$ for using between two vectors represents the \emph{outer product} in linear algebra which gives rise to a matrix.} to emphasize the tensor product structure of $\mathcal{H}_{\Sigma}$\cite{1508.02595}. Then density matrix is
\begin{equation}
\rho_{\Sigma} = |\psi\rangle_{\Sigma}\langle\psi|_{\Sigma}=\sum\limits_{ijkl}c_{ij}c^{*}_{kl}|i\rangle_{A}|j\rangle_{B}\langle{k}|_{A}\langle{l}|_{B}.
\label{}
\end{equation}
Now suppose $\mathcal{O}_{A}$ is an observable for the sub-system $A$ and for the composite system $\Sigma$ the observable can be regarded as a \emph{matrix tensor product} (Kcroneker product) $\mathcal{O}=\mathcal{O}_{A}\otimes\mathbbm{1}_{B}$ and $\langle\mathcal{O}\rangle=\text{tr}_{\Sigma}(\rho_{\Sigma}\mathcal{O})$. In the disentangled representation,
\begin{equation}
\begin{split}
\langle\mathcal{O}\rangle &= \langle\psi|\mathcal{O}_{A}\otimes\mathbbm{1}_{B}|\psi\rangle \\
&= \sum\limits_{ijkl}c^{*}_{kl}c_{ij}\langle{k}|_{A}\langle{l}|_{B}\mathcal{O}_{A}\otimes\mathbbm{1}_{B}|i\rangle_{A}|j\rangle_{B}\\
&= \sum\limits_{ijkl}c^{*}_{kl}c_{ij}\delta_{lj}\langle{k}|_{A}\mathcal{O}_{A}|i\rangle_{A} \\
&= \sum\limits_{ijk}c_{ij}c^{*}_{kj}\langle{k}|_{A}\mathcal{O}_{A}|i\rangle_{A}.
\end{split}
\label{eq:01}
\end{equation}
On the other hand,
\begin{equation}
\begin{split}
\text{tr}_{B}\rho_{\Sigma} &= \sum\limits_{l'}\langle{l'}|_{B}\sum\limits_{ijkl}c_{ij}c^{*}_{kl}|i\rangle_{A}|j\rangle_{B}\langle{k}|_{A}\langle{l}|_{B}|l'\rangle_{B} \\
&= \sum\limits_{ijk}c_{ij}c^{*}_{kj}|i\rangle_{A}\langle{k}|_{A},
\label{}
\end{split}
\end{equation}
and
\begin{equation}
\begin{split}
\text{tr}_{A}(\rho_{A}\mathcal{O}_{A}) &= \sum\limits_{k'}\langle{k'}|_{A}\sum\limits_{ijk}c_{ij}c^{*}_{kj}|i\rangle_{A}\langle{k}|_{A}\mathcal{O}_{A}|k'\rangle_{A} \\
&= \sum\limits_{ijk}c_{ij}c^{*}_{kj}\langle{k}|_{A}\mathcal{O}_{A}|i\rangle_{A}.
\label{eq:02}
\end{split}
\end{equation}
According to Eq. \ref{eq:01} and \ref{eq:02}, $\langle\mathcal{O}\rangle = \text{tr}_{A}(\rho_{A}\mathcal{O}_{A})$ and $\rho_{A} = \sum_{ijk}c_{ij}c^{*}_{kj}|i\rangle_{A}\langle{k}|_{A}\equiv\sum_{j}|j\rangle\langle{j}|$, where $|j\rangle\equiv\sum_{i}c_{ij}|i\rangle_{A}$ is defined in the Hilbert space $\mathcal{H}_{A}$ but not normalized. Note that
\begin{itemize}
\item If the state $|\psi\rangle$ for the whole system is a pure state, the sub-system $A$ is not expected to be in a pure state. $\rho_{A}^{2}\neq\rho_{A}$ if sub-system $A$ is in a mixed state. $\rho_{A}$ also can be viewed as the reduced state from $|\psi\rangle$ and the rest sub-system $B$ is regarded as a coupled auxiliary system.
\end{itemize}

\subsection{Schmidt decomposition and entanglement entropy}
Singular value decomposition (SVD) theorem manifests that for a rectangular matrix $\mathbf{C}\in\mathbbm{C}^{m\times{n}}$, it can be decomposed into the form
\begin{equation}
\mathbf{C} = \mathbf{U}\mathbf{\Lambda}\mathbf{V}^{\dagger},
\label{}
\end{equation}
where $\mathbf{U}$ and $\mathbf{V}$ are complex unitary matrices and $\mathbf{\Lambda}\in\mathbbm{R}^{m\times{n}}$ is diagonal with non-zero diagonal singular values $\{\lambda_{0},\cdots,\lambda_{r-1}\}$, $\text{rank}(\mathbf{C}) = r$. In the direct product representation $\{(|i\rangle_{A}, |j\rangle_{B})\}$, the state $|\psi\rangle$ is represented by the matrix $(c_{ij})$
\begin{widetext}
\begin{equation}
|\psi\rangle_{\Sigma} = (|0\rangle_{A}, |1\rangle_{A}, \cdots, |d_{A}-1\rangle_{A})
\begin{pmatrix}
c_{00} & c_{01} & \cdots & c_{0, d_{B}-1} \\
c_{10} & c_{11} & \cdots & c_{1, d_{B}-1} \\
\vdots & \vdots & \ddots & \vdots \\
c_{d_{A}-1, 0} & c_{d_{A}-1, 1} & \cdots & c_{d_{A}-1, d_{B}-1}
\end{pmatrix}
\begin{pmatrix}
|0\rangle_{B} \\
|1\rangle_{B} \\
\vdots \\
|d_{B}-1\rangle_{B}
\end{pmatrix}.
\label{}
\end{equation}
\end{widetext}
Note that the Cartesian direct product means \emph{ordered pair} and a disentangled directly product Cartesian term means there is absolutely no information of one we can retrieve from the other and the combination of them precisely describes a basis state vector in $\mathcal{H}_{\Sigma}$. If we carry out the unitary rotations $(|0\rangle_{A}, |1\rangle_{A}, \cdots, |d_{A}-1\rangle_{A})\mathbf{U}^{\dagger}$ and $\mathbf{V}(|0\rangle_{B}, |1\rangle_{B}, \cdots, |d_{B}-1\rangle_{B})^{T}$,
\begin{equation}
|\psi\rangle_{\Sigma} = \sum\limits_{k = 0}^{r-1}\lambda_{k}|k\rangle_{A}|k\rangle_{B}.
\label{}
\end{equation}
If $|\psi\rangle$ is normalized, $\sum_{k = 0}^{r-1}|\lambda_{k}|^{2} = 1$. In this sense, the von Neumann entropy is defined as the EE for sub-system $A$
\begin{equation}
S_{A} = -\text{tr}(\rho_{A}\ln\rho_{A}) = -\sum\limits_{k = 0}^{r-1}|\lambda_{k}|^{2}\ln|\lambda_{k}|^{2} = -\text{tr}(\rho_{B}\ln\rho_{B}). 
\label{}
\end{equation}
The EE is symmetric if the total system $\Sigma$ is in a pure state. Also note the rank of the representation matrix $\mathbf{C}$ is intrinsic and important. It is independent of specific representations thus the entanglement entropy have nothing to do with representations. Specially, if there is just one none zero singular value $\lambda_{0} = 1$ we find that $S_{A} = S_{B} = 0$. It is indeed a measure of the entanglement between two sub-systems. In this sense, we can also define what is a entangled state precisely.

\emph{If the rank of the representation matrix in the direct product representation is larger than 1, the pure state represented by this very matrix is said to be entangled. Otherwise, it is a disentangled, direct product state.}

In this sense, seemingly entanglement is more intrinsic than the superposition principle in quantum mechanics as it concerns about the \emph{rank} of the representation matrix rather than specific representations. Rank of a matrix is invariant under similarity transformations in different representations.

In a word, the above discussion implies that if we would like to investigate the entanglement between two sub-systems of a whole composite system $\Sigma$, the practical procedure is
\begin{itemize}
\item The first and most pivotal step is to expand the state $|\psi\rangle_{\Sigma}$ in terms of a basis written in a direct product of the sub-systems $A$ and $B$.
\item Then trace out one of the sub-systems to obtain the reduced density matrix.
\end{itemize}

\bibliographystyle{apsrev4-1} 
\bibliography{tjchain-phasestring-mutual-entanglement}
\end{document}